\documentclass[aps,prl,twocolumn,floatfix]{revtex4-2}
\usepackage{graphicx}
\usepackage{amsmath,amssymb}
\usepackage{bm}
\usepackage{xcolor}
\usepackage{txfonts}
\usepackage[colorlinks=true, linkcolor=blue, citecolor=blue, urlcolor=blue]{hyperref}
\usepackage{dsfont}
\newcommand{\sectionprl}[1]{{\em #1}\/---}

\begin{document}

\title{Relaxation of Sliding Friction from a Statistical Model of Aging Contacts}

\author{Ryudo Suzuki}
\affiliation{Department of Physics, Kyoto University, Kyoto 606-8502, Japan}

\date{\today}

\begin{abstract}
Frictional relaxation after a sudden velocity change underlies
the phenomenological laws that describe the stability of steady 
sliding and of seismic faults. To connect this relaxation 
with measurable statistics of microscopic contacts, we introduce 
a minimal model incorporating logarithmic aging of contact forces 
and a power-law contact-size distribution reflecting fractal 
interface roughness, and derive the relaxation function analytically. 
Depending on the power-law exponent $\alpha$, the intermediate-time 
response exhibits a plateau, logarithmic decay, or power-law decay, 
in contrast to the exponential decay assumed in the phenomenological 
laws. The theory provides experimentally 
testable links between microscopic contact statistics and 
macroscopic frictional relaxation.
\end{abstract}

\maketitle

\sectionprl{Introduction.}
Relaxation in response to an external perturbation provides
fundamental information about a system and has been studied across
a wide range of fields, including structural relaxation in glasses
\cite{BerthierBiroli2011}, dielectric relaxation
\cite{KremerSchonhals2003}, and stress relaxation in rheology
\cite{Ferry1980}.
In sliding friction, velocity-step experiments, which measure the
relaxation of the frictional force after a sudden change in the
sliding velocity, have been widely performed
\cite{Dieterich1979,Dieterich1981,Ruina1983,Marone1998}.
This relaxation is conventionally described as an exponential decay
within phenomenological rate-and-state friction laws
(e.g., the slip law \cite{Ruina1983,Marone1998}), which underlie 
analyses of the stability of steady sliding \cite{RiceRuina1983,Heslot1994} and of
seismic faults \cite{Marone1998,Scholz1998,Scholz2019}.
Microscopic interpretations of these laws have also been proposed
\cite{Dieterich1994,Marone1998,BaumbergerCaroli2006,Putelat2011}, 
and a natural next step is to identify how the relaxation, 
including its shape and characteristic scales \cite{Hatano2015}, 
emerges from the microscopic properties of the interface.

\begin{figure}[!tbp]
    \centering
    \includegraphics[width=\columnwidth]{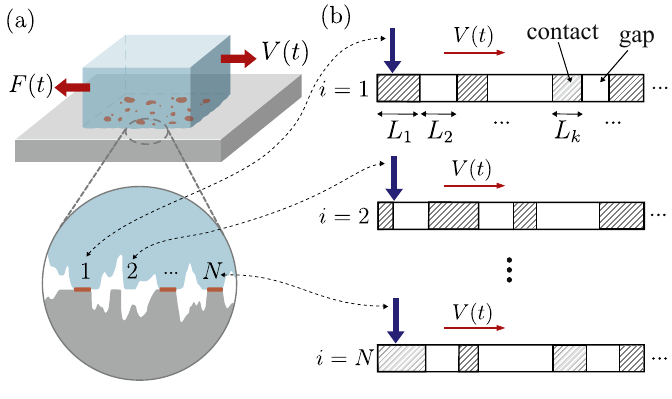}
    \caption{
        Schematic of the model.
        (a) Frictional interface. A block (blue) slides at velocity $V(t)$
        on a rough substrate (gray) and experiences a frictional force $F(t)$.
        The two surfaces touch through numerous microscopic contact
        junctions (red dots).
        Magnified view: roughness peaks $i=1,2,\cdots,N$ 
        on the block form microcontacts (red) with roughness peaks 
        on the substrate.
        (b) One-dimensional model of this study, obtained by simplifying (a).
        Each roughness peak $i$ on the block is represented 
        by an arrow (blue) that moves along a rail at velocity $V(t)$.
        On the rail, contact regions (contact, hatched) and gap regions
        (gap, white) alternate, and their lengths $L_k$ are sampled
        independently from the distributions $\rho_{\mathrm{c}}(L)$ and
        $\rho_{\mathrm{g}}(L)$, respectively.
        A rail is generated independently for each arrow, and a contact
        force acts on an arrow when it is located on a contact region.
    }
    \label{fig1}
\end{figure}

Naturally formed frictional interfaces 
exhibit fractal roughness over a wide range of length scales 
\cite{persson2014}.
Consequently, macroscopic solids touch each other through numerous
microscopic contact junctions [see Fig.~\ref{fig1}(a)], 
and the macroscopic frictional force arises as 
the sum of the adhesive forces acting at these junctions 
\cite{BowdenTabor2001}.
Reflecting the fractal nature of the interface, the microcontact-size 
distribution follows a power law
\cite{Dieterich1996,MuserWang2018}.
Moreover, the adhesive force at each junction exhibits \textit{aging}, 
growing logarithmically with the contact time 
\cite{Dieterich1996,Berthoud1999,Li2011}.
During sliding, new microcontacts repeatedly form, 
strengthen through aging while they remain in contact, 
and eventually rupture
\cite{Dieterich1994,Berthoud1999,Li2011,Marone1998,BaumbergerCaroli2006}.

Within this microscopic picture, the relaxation after a velocity step
can be understood as the replacement of contacts formed before the 
velocity change by a new population corresponding to the new sliding 
velocity \cite{Dieterich1994,Marone1998,BaumbergerCaroli2006}.
For example, after an increase in the sliding velocity, long-lived and
strong contacts are progressively replaced by shorter-lived and weaker
ones, causing the frictional force to relax toward its new steady-state 
value.

In this Letter, we introduce a minimal model combining logarithmic
aging of contact forces with a power-law distribution of contact
lengths, and analytically derive the relaxation of the frictional force.
The resulting relaxation deviates from the exponential decay assumed
in conventional rate-and-state laws:
as shown in Eq.~\eqref{eq:relax_scaling},
in an intermediate-time regime it exhibits a plateau,
logarithmic decay, or power-law decay, depending on the
exponent $\alpha$ of the contact-length distribution.
The form of the relaxation is thus set by the contact statistics
associated with the fractal nature of the interface.
Our results provide a relation linking macroscopic friction to
measurable microcontact statistics, which can be tested directly by
combining velocity-step measurements with interface imaging
\cite{Sahli2018,Gvirtzman2025,Bennett2017}.

\sectionprl{Model.}
We introduce a model for the relaxation of the
frictional force following an instantaneous change in the sliding 
velocity, incorporating a power-law contact-size distribution 
together with contact aging and renewal.
At an interface between two rough solid surfaces 
[see Fig.~\ref{fig1}(a)], numerous microcontacts repeatedly 
form and rupture as the surfaces slide past each other.
To simplify this situation, as shown in Fig.~\ref{fig1}(b),
we consider $N$ mutually independent arrows moving at velocity 
$V(t)$ along one-dimensional rails.
Each arrow represents a roughness peak on a surface 
and experiences a contact force $f_i\,(i=1,\cdots,N)$.
Spatial correlations and interactions between contacts are neglected.
The frictional force $F(t)$ of the  system is defined as the
average of the contact forces acting on the arrows:
\begin{align}\label{eq:force_average}
    F(t) = \frac{1}{N} \sum_{i=1}^N f_i(t).
\end{align}
Since we focus on the shape of the relaxation function after a velocity
step, the normalization is chosen for later convenience.

Next, we specify the contact force acting on each arrow.
Contact and gap regions alternate along each one-dimensional rail.
When arrow $i$ is on a contact region, the contact force is given by
\begin{align}
    f_i(t) = \bar{f}\!\left(V(t)\right) Z\!\left(\theta_i(t)\right),
\end{align}
whereas $f_i(t)=0$ when it is on a gap region.
Here, $\bar{f}(V)$ represents the velocity-dependent contribution 
to the contact force and gives rise to the instantaneous response 
following a velocity step.
The variable $\theta_i$ is the contact time elapsed since 
the arrow entered the current contact region, 
and it grows from zero each time the arrow enters a contact region.
The function $Z(\theta)$ describes the aging of the contact force.
At microcontacts, the contact area \cite{Dieterich1994} and 
bonding strength \cite{Li2011,Liu2012} are known to increase 
logarithmically with the contact time.
We therefore assume the aging function
\begin{align}\label{eq:aging}
Z(\theta)
=
1+c\log\!\left(1+\frac{\theta}{\tau}\right),
\end{align}
where $\tau$ is the time scale of the aging 
and $c$ represents its strength.
Hereafter, we set $\bar f(V)\equiv f_0$ for simplicity, 
since its velocity dependence does not affect the form of the 
relaxation. With this choice, we neglect the instantaneous change 
in the friction force following a velocity step.

The lengths $L$ of the contact and gap regions are independently 
sampled from distributions $\rho_{\mathrm{c}}(L)$ and 
$\rho_{\mathrm{g}}(L)$, respectively.
We denote the corresponding mean lengths by
$\langle L \rangle_a
\equiv \int_0^{\infty} L\, \rho_a(L)\, dL$, $a\in \{\mathrm{c},\mathrm{g}\}$.
The gap-length distribution $\rho_{\mathrm{g}}(L)$ is arbitrary except
that its mean $\langle L \rangle_{\mathrm{g}}$ is finite.
Reflecting the fractal nature of rough solid interfaces,
we assume that the contact-length distribution follows 
a truncated power law with exponent $\alpha>1$: 
\begin{align}\label{eq:size_dist}
    \rho_{\mathrm{c}}(L)
    =
    C L^{-\alpha}
    \mathds{1}_{[L_{\min},L_{\max}]}(L),
    \quad
    C\equiv \frac{\alpha-1}{L_{\min}^{1-\alpha}-L_{\max}^{1-\alpha}},
\end{align}
where $\mathds{1}_{[L_{\min},L_{\max}]}(L)$ is the indicator function,
which equals unity for $L\in[L_{\min},L_{\max}]$ and zero otherwise.
For real rough interfaces, the range of wave numbers 
over which the surface power spectrum follows a power law is 
bounded by microscopic and macroscopic cutoffs \cite{persson2014}.
Correspondingly, we introduce a lower cutoff $L_{\min}$ and 
an upper cutoff $L_{\max}$ for the contact-length distribution.
In this Letter, we assume $L_{\min}\sim1\,\mathrm{nm}$ and
$L_{\max}\sim100\,\mu\mathrm{m}$ as typical values 
\cite{Dieterich1996,persson2014} and consider 
the regime $L_{\min}\ll L_{\max}$.
As shown below, $\rho_{\mathrm{g}}(L)$ does not affect the functional 
form of the relaxation. We therefore set
$\rho_{\mathrm{g}}(L)=\rho_{\mathrm{c}}(L)$ in the numerical
calculations.

In this model, the sliding velocity $V(t)$ is externally prescribed.
We consider a velocity-step protocol in which, 
after the system reaches a steady state at velocity $V_1$, 
the velocity is instantaneously changed to $V_2$ at time $t=0$:
\begin{align}\label{eq:velocity_step}
    V(t) =
    \begin{cases}
        V_1 & t<0,\\
        V_2 & t\geq 0.
    \end{cases}
\end{align}

\begin{figure}[!tbp]
    \centering
    \includegraphics[width=\columnwidth]{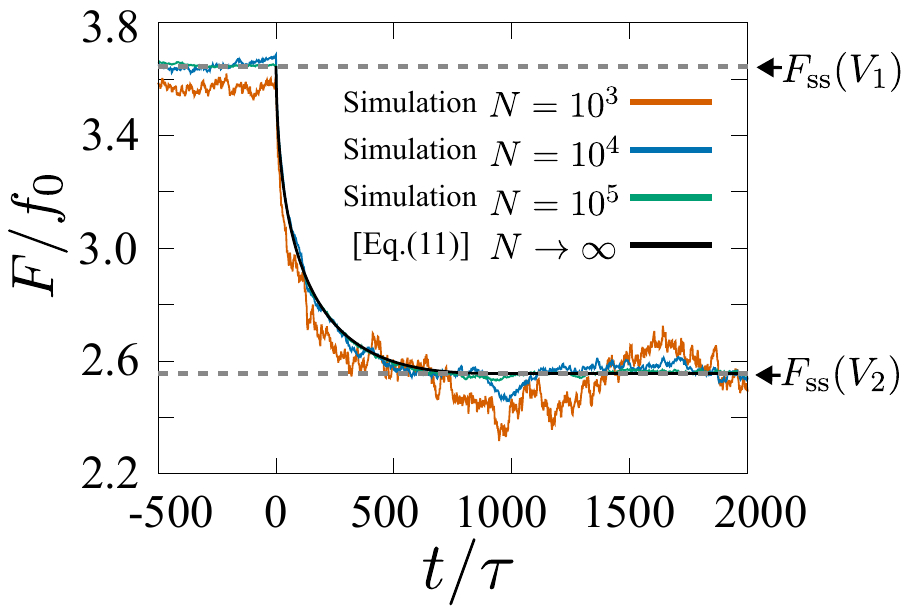}
    \caption{
        Relaxation of the frictional force $F(t)$ after the velocity
        step.
        The numerical results of the microscopic model with a finite
        number of arrows $N$ converge, as $N$ increases, to the
        analytic result \eqref{eq:Ninfty} obtained in the
        $N\to\infty$ limit (black solid line).
        The parameters are $r=V_1/V_2=0.1$, $\alpha=1.5$,
        $\xi_{\min}=10^{-2}$, $\xi_{\max}=10^3$, $f_0=1$, and $c=1$.
    }
    \label{fig2}
\end{figure}

Apart from $\alpha$ and $c$, the model is characterized 
by three dimensionless parameters: $\xi_{\min}=L_{\min}/(V_2 \tau)$, 
$\xi_{\max}=L_{\max}/(V_2 \tau)$, and the velocity ratio $r=V_1/V_2$.
In the numerical calculations, we use $\xi_{\min}=10^{-2}$,
$\xi_{\max}\in [10^0,10^4]$, and $r=0.1$, 
so that $\xi_{\min}\ll \xi_{\max}$.
In typical velocity-step experiments, the sliding velocity is of the
order of a few micrometers per second. Taking
$V_2=1\,\mu\mathrm{m}\,\mathrm{s}^{-1}$ and
$\tau=0.1\,\mathrm{s}$ \cite{Li2011,Berthoud1999},
these parameters correspond to
$V_1=0.1\,\mu\mathrm{m}\,\mathrm{s}^{-1}$,
$L_{\min}=1\,\mathrm{nm}$, and
$L_{\max}=0.1\,\mu\mathrm{m}$--$1\,\mathrm{mm}$.

Figure~\ref{fig2} shows the frictional relaxation after the velocity
step for different $N$. The results converge to a deterministic curve
as $N$ increases.
In the following, we consider the $N\to\infty$ limit to
investigate the functional form of the relaxation.

\sectionprl{Large-$N$ limit.}
We now analyze the relaxation process of the microscopic model
in the $N\to\infty$ limit.
In this limit, the frictional force can be expressed in terms of
the contact-time distribution $P_t(\theta)$:
\begin{align}\label{eq:F_Pt}
    F(t) =
    f_0\, \Phi \int_0^\infty Z(\theta) P_t(\theta)\, d\theta,
    \quad
    \Phi \equiv
    \frac{\langle L\rangle_{\mathrm{c}}}{\langle L\rangle_{\mathrm{c}}
    + \langle L\rangle_{\mathrm{g}}}.
\end{align}
Here, $\Phi$ is the probability that an arbitrary arrow is in
a contact region, and $P_t(\theta)$ is the probability density
of the contact age $\theta$ conditioned on an arrow being in
contact. Since every arrow in contact has some age
$\theta\in[0,\infty)$, the distribution is normalized as
$\int_0^\infty P_t(\theta)\,d\theta=1$ by definition.

To derive the frictional relaxation, we consider the time 
evolution of the contact-time distribution $P_t(\theta)$ 
under a general velocity $V(t)$. A contact of age $\theta$
at time $t$ was formed at time $t-\theta$ and has traveled 
the distance $x(t,\theta)=\int_{t-\theta}^{t}V(s)\,ds$.
As long as the contact survives, its age increases at unit 
rate, $d\theta/dt=1$, giving rise to the
advection term $\partial P_t(\theta)/\partial\theta$ below.
Its survival probability after traveling a distance 
$x$ is $S(x)=\int_x^\infty \rho_{\mathrm c}(L)\,dL$, 
and the corresponding hazard rate per unit distance is
$\lambda(x)\equiv\rho_{\mathrm c}(x)/S(x)$.
Therefore, a contact of age $\theta$ is lost at 
the rate $V(t)\lambda(x(t,\theta))$.
Introducing the influx $J(t)$ of newly formed contacts at
$\theta=0$, we obtain
\begin{align}\label{eq:master_eq}
    \frac{\partial P_t(\theta)}{\partial t}
    +\frac{\partial P_t(\theta)}{\partial\theta}
    =
    -V(t)\lambda(x(t,\theta))P_t(\theta)
    +J(t)\delta(\theta).
\end{align}
Probability conservation requires the influx to equal the total
contact-loss rate:
\begin{align}\label{eq:flux}
    J(t)
    =
    V(t)\int_0^\infty
    \lambda(x(t,\theta))P_t(\theta)\,d\theta.
\end{align}
Integrating Eq.~\eqref{eq:master_eq} across $\theta=0$
gives the boundary condition $P_t(0)=J(t)$.

\begin{figure}[!tbp]
    \centering
    \includegraphics[width=0.9\columnwidth]{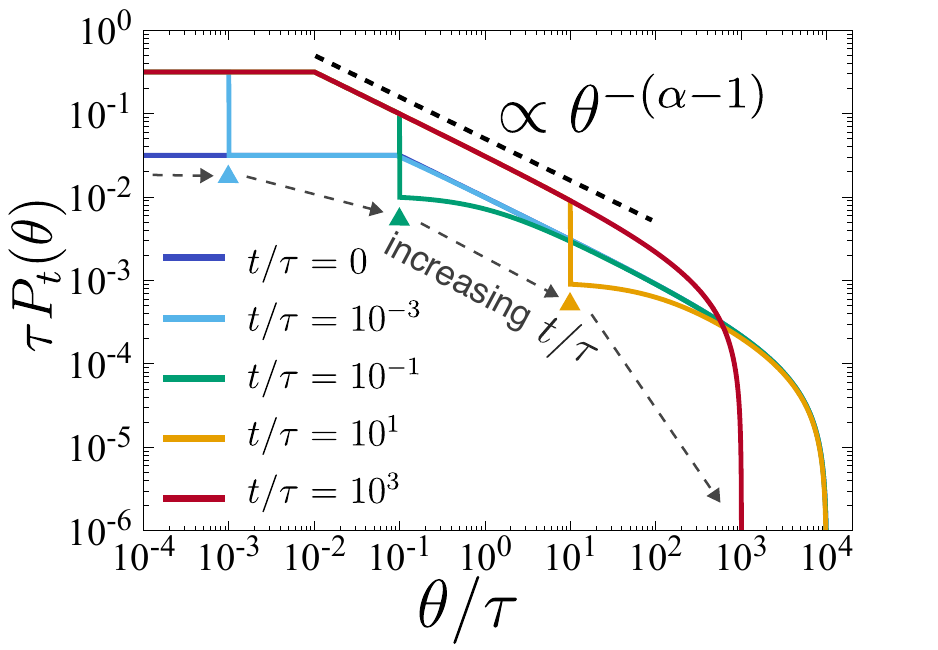}
    \caption{
        Time evolution of the contact-time distribution $P_t(\theta)$
        after the velocity step.
        The color of each curve represents the elapsed time $t/\tau$
        after the velocity change.
        As time proceeds, the discontinuity at $\theta=t$ moves 
        to the right, as indicated by the arrows,
        and the old contact population is 
        replaced by the new one.
        In the intermediate range of $\theta$, the distribution follows
        the power law $P_t(\theta)\propto\theta^{-(\alpha-1)}$
        (dashed line), which originates from the power-law decay 
        of the survival function $S(x)$. See Appendix~A in the 
        End Matter for the derivation.
        The parameters are the same as in Fig.~\ref{fig2}.
    }
    \label{fig3}
\end{figure}

Solving the master equation \eqref{eq:master_eq} with the boundary
condition $P_t(0)=J(t)$, we obtain
$P_t(\theta)=J(t-\theta)S(x(t,\theta))$.
This expression means that an arrow that entered a contact region 
at time $t-\theta$ remains in the same contact region after 
traveling the distance $x(t,\theta)$ with probability 
$S(x(t,\theta))$.
Substituting this solution into Eq.~\eqref{eq:flux} yields
$J(t)=V(t)/\langle L\rangle_{\mathrm{c}}$.
The solution of the master equation is therefore
\begin{align}\label{eq:Pt_sol}
    P_t(\theta)
    =
    \frac{V(t-\theta)S(x(t,\theta))}{\langle L\rangle_{\mathrm{c}}}.
\end{align}
See Supplemental Material \cite{SM} for the derivation of 
Eq.~\eqref{eq:Pt_sol}. Accordingly, after the velocity step in 
Eq.~\eqref{eq:velocity_step}, the contact-time distribution is
\begin{align}\label{eq:Pt_step}
    P_t(\theta)
    =\frac{1}{\langle L\rangle_{\mathrm{c}}}
    \begin{cases}
        V_2 S(V_2\theta),
        & \theta<t,\\
        V_1 S\!\left(V_1(\theta-t)+V_2t\right),
        & \theta>t.
    \end{cases}
\end{align}
Figure~\ref{fig3} shows the time evolution of the contact-time
distribution after the velocity step.
The region $\theta<t$ corresponds to new contacts formed after the
velocity step, while $\theta>t$ corresponds to old contacts remaining
from before the step.
As time proceeds, the boundary $\theta=t$ moves to the right, and the
old contact population is progressively replaced by the new one.

\begin{figure*}[!t]
    \centering
    \includegraphics[width=0.93\textwidth]{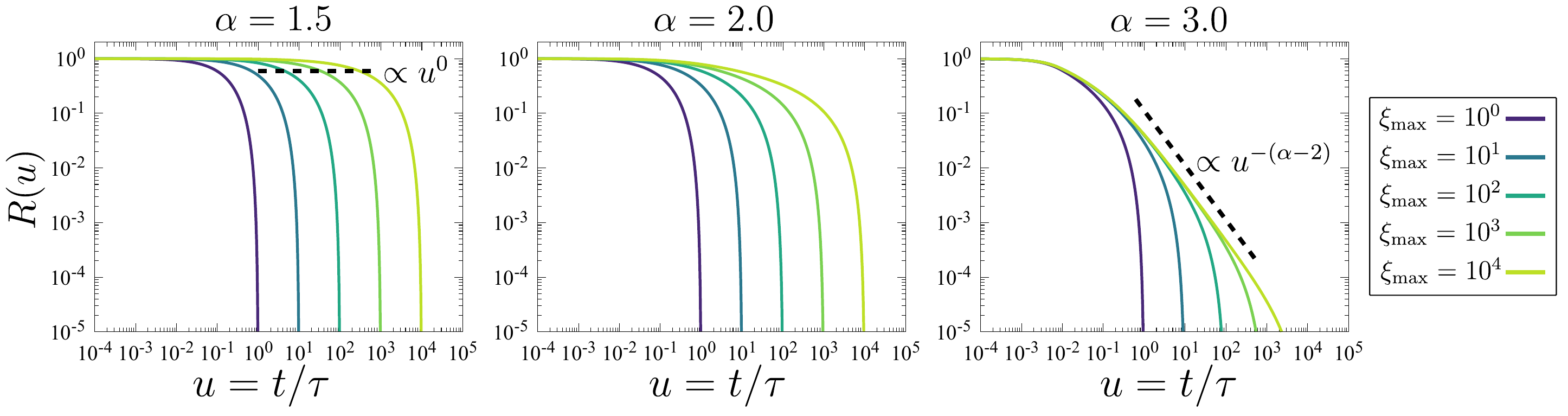}
    \caption{
    Time evolution of the relaxation function $R(u)$, where $u=t/\tau$
    is the dimensionless time.
    The panels show the cases $\alpha=1.5,\,2.0,\,3.0$, respectively,
    and the color of each curve represents the value of the macroscopic
    cutoff $\xi_{\max}$ ($\xi_{\max}=10^{0}$--$10^{4}$).
    In the intermediate-time regime, a plateau, logarithmic decay, or
    power-law decay appears depending on $\alpha$, as shown in
    Eq.~\eqref{eq:relax_scaling}.
    This intermediate-time regime widens as $\xi_{\max}$ increases.
    The parameters are the same as in Fig.~\ref{fig2}.
    }
    \label{fig4}
\end{figure*}

Substituting Eq.~\eqref{eq:Pt_step} into Eq.~\eqref{eq:F_Pt},
we obtain the frictional force for $t>0$:
\begin{align}\label{eq:Ninfty}
    F(t) = \frac{f_0\Phi}{\langle L\rangle_{\mathrm{c}}}
    \Bigg[
        &\int_0^t d\theta\, Z(\theta)\,V_2\,S(V_2\theta) \notag\\
        &+ \int_t^{\infty} d\theta\, Z(\theta)\,V_1\,
        S\!\left(V_1(\theta-t) + V_2 t\right)
    \Bigg].
\end{align}
The two terms correspond to the new ($\theta<t$) and old ($\theta>t$)
contact populations in Eq.~\eqref{eq:Pt_step}, respectively.
The corresponding steady-state friction force at sliding
velocity $V$ is 
\begin{align}\label{eq:Fss}
    F_{\mathrm{ss}}(V) = \frac{f_0 \Phi V}{\langle L \rangle_{\mathrm{c}}} \int_0^\infty d\theta\, Z(\theta)S(V \theta),
\end{align}
whose velocity dependence is given in Appendix~B.

Equation~ \eqref{eq:Ninfty} describes the relaxation of the friction
force from the initial steady-state value $F_{\mathrm{ss}}(V_1)$ to the
new steady-state value $F_{\mathrm{ss}}(V_2)$.
As shown in Fig.~\ref{fig2}, the simulation results of the microscopic
model with finite $N$ converge to the analytic result in 
Eq.~\eqref{eq:Ninfty}.
In the following, we use this expression to determine 
the functional form of the relaxation.

\sectionprl{Relaxation function.}
To investigate the relaxation of the frictional force after the
velocity step, we introduce the dimensionless time $u\equiv t/\tau$ 
and define the normalized relaxation function
\begin{align}\label{eq:Rdef}
    R(u) \equiv 
    \frac{F(\tau u) - F_{\mathrm{ss}}(V_2)}{F_{\mathrm{ss}}(V_1) 
    - F_{\mathrm{ss}}(V_2)},
\end{align}
which satisfies $R(0)=1$ and decays to zero as the friction force
approaches the new steady state.

Substituting Eqs.~\eqref{eq:Ninfty} and \eqref{eq:Fss} into
Eq.~\eqref{eq:Rdef} and changing variables, we obtain
\begin{equation}\label{eq:R_express}
    R(u)=\frac{\mathcal{N}(u)}{\mathcal{N}(0)},
\end{equation}
where
\begin{equation}\label{eq:N_def}
    \mathcal{N}(u)
    \equiv \int_u^\infty d\xi'\, s(\xi')
    \left[
        z\!\left(u+r^{-1}(\xi'-u)\right) - z(\xi')
    \right],
\end{equation}
with $s(y)\equiv S(V_2\tau y)$ and $z(y)\equiv Z(\tau y)$.
Notably, the gap-length distribution $\rho_{\mathrm{g}}(L)$ enters
Eqs.~\eqref{eq:Ninfty} and \eqref{eq:Fss} only through $\Phi$ and
hence cancels in $R(u)$. Therefore, the shape of the relaxation 
is independent of the gap statistics and is controlled by the 
contact-length distribution $\rho_{\rm c}(L)$.

Equation~\eqref{eq:N_def} shows that only old contacts surviving beyond
the dimensionless slip distance $u$, namely those with $\xi'>u$,
contribute to $\mathcal{N}(u)$.
Such a contact traveled a distance $\xi'-u$ at velocity $V_1$ before
the step and a distance $u$ at velocity $V_2$ after it, so that its
dimensionless age is $u+r^{-1}(\xi'-u)$ and its aging contribution
is $z(u+r^{-1}(\xi'-u))$.
By contrast, $z(\xi')$ is the aging contribution that the same
contact would have accumulated over the slip distance $\xi'$ at
velocity $V_2$ alone.
The integrand of Eq.~\eqref{eq:N_def} is the difference between the
two, weighted by the survival probability $s(\xi')$. Thus,
$\mathcal{N}(u)$ quantifies the excess aging caused by the
velocity history, and $\mathcal{N}(0)$ normalizes it by its 
initial value.

To determine the functional form of $R(u)$, we evaluate
Eq.~\eqref{eq:R_express} in the intermediate-time regime
$\max\{1,\xi_{\min}\}\ll u \ll \xi_{\max}$.
Substituting Eqs.~\eqref{eq:aging} and \eqref{eq:size_dist},
changing variables as $\xi'=ux$, and expanding for $u\gg1$,
we obtain
\begin{equation}\label{eq:R_asymptotic}
    \begin{gathered}
        R(u)
        \simeq
        \frac{c\,u^{2-\alpha}}{A\,\mathcal{N}(0)}\,
        I_\alpha\!\left(\frac{\xi_{\max}}{u}\right),
        \,\,
        I_\alpha(\Lambda)
        \equiv
        \int_1^\Lambda
        \left(x^{1-\alpha}-\Lambda^{1-\alpha}\right)
        g(x)\,dx,
    \end{gathered}
\end{equation}
where $A\equiv \xi_{\min}^{1-\alpha}-\xi_{\max}^{1-\alpha}$ and
$g(x)\equiv\log\left[(1+(x-1)/r)/x\right]$.
See Supplemental Material \cite{SM} for the derivation.

In this regime, $\Lambda=\xi_{\max}/u\gg1$.
Evaluating $I_\alpha(\Lambda)$ for $\Lambda\gg1$ and
$\mathcal{N}(0)$ for $\xi_{\max}\gg\max\{\xi_{\min},1,r\}$,
we find
\begin{equation}\label{eq:relax_scaling}
    R(u) \simeq
    \begin{cases}
        1, & 1<\alpha<2,\\
        1 - \log u/\log\xi_{\max}, & \alpha=2,\\
        \mathcal{C}_\alpha\, u^{-(\alpha-2)}, & \alpha>2,
    \end{cases}
\end{equation}
where $\mathcal{C}_\alpha$ is an $O(1)$ amplitude that depends 
on the velocity ratio $r$ and $\xi_{\min}$.
See Supplemental Material \cite{SM} for its explicit form.
These asymptotic forms agree with the numerical results in
Fig.~\ref{fig4}.
Outside this regime, $R(u)$ starts from a plateau at
$u\ll\min\{1,\xi_{\min}\}$ and vanishes as $(\xi_{\max}-u)^3$ as
$u\to\xi_{\max}$, where the relaxation completes (Appendix~C).

The functional form of the relaxation in the intermediate-time
regime is thus governed by the exponent $\alpha$ of the
contact-length distribution: a plateau for $\alpha<2$, power-law
decay for $\alpha>2$, and logarithmic decay at the boundary
$\alpha=2$.
Physically, decreasing $\alpha$ shifts the weight of the
distribution toward long contacts near the upper cutoff $L_{\max}$,
whose slow renewal maintains the plateau, whereas increasing
$\alpha$ shifts it toward short contacts near the lower cutoff
$L_{\min}$, whose rapid renewal speeds up the relaxation
(Appendix~D).

\sectionprl{Discussion.}
In this Letter, we introduced a minimal model to clarify how 
the functional form of frictional relaxation is determined by 
the statistics of microscopic contacts.
We showed that the resulting relaxation deviates from a simple 
exponential decay: depending on the exponent $\alpha$, 
it exhibits a plateau, logarithmic decay, or 
power-law decay [Eq.~\eqref{eq:relax_scaling}], 
reflecting the contact statistics of the fractal 
interface.

Among the rate-and-state friction laws widely used in geoscience 
and other fields \cite{Dieterich1979,Ruina1983,Marone1998}, 
the slip law describes the relaxation after a velocity step as 
an exponential decay \cite{Ruina1983}. Experimentally, however, 
it has not been established that the relaxation is strictly 
exponential of follows other forms 
\cite{Kilgore1993,Bhattacharya2015,Bhattacharya2022}, 
motivating tests of the distinct forms predicted by
Eq.~\eqref{eq:relax_scaling}.

Finally, our results establish an experimentally testable link between
microscopically measurable quantities and the macroscopic
frictional relaxation. This relation can be tested directly 
in systems where microcontacts are imaged 
through transparent materials
\cite{Sahli2018,Gvirtzman2025,Bennett2017}.
First, during steady sliding, one measures the intermediate 
power-law regime $P_t(\theta)\propto \theta^{-(\alpha-1)}$ of 
the contact-time distribution and estimates $\alpha$.
Then, by performing a velocity-step experiment,
one can test whether the measured relaxation function follows the
prediction in Eq.~\eqref{eq:relax_scaling}.
Such measurements would provide a direct 
test of the microscopic origin of phenomenological friction laws.

\sectionprl{Acknowledgments.}
The author thanks S.-i.~Sasa and S.~Poincloux for their critical 
reading of the manuscript and valuable comments, T.~Hatano for 
an intensive lecture course that inspired this work,  
J.-C.~Delvenne and Y.~Yanagisawa for helpful discussions, 
and K.~Yamamoto for drawing Fig.~1. 
OpenAI ChatGPT (GPT-5.6 Sol) and Anthropic Claude (Fable 5) 
were used to assist with numerical code development, 
theoretical analysis, and manuscript editing.
The author reviewed all AI-assisted outputs, independently
verified the theoretical analysis, 
and takes full responsibility for the content.
This work was supported by JST SPRING, Grant No.~JPMJSP2110.

\sectionprl{Data Availability.}
The numerical data and source code used to generate Figs.~2--6
are openly available in Ref.~\cite{SuzukiData2026}.

\makeatletter
\let\auto@bib@innerbib\@empty
\makeatother

\newpage

\onecolumngrid
\vspace{5mm}
\begin{center}
    \large \textbf{End Matter}
\end{center}
\twocolumngrid

\bigskip
\sectionprl{Appendix A: Power-law regime of $P_t(\theta)$}
\renewcommand{\theequation}{A\arabic{equation}}
\renewcommand{\theHequation}{A.\arabic{equation}}
\setcounter{equation}{0}
For the truncated power law in Eq.~\eqref{eq:size_dist},
the survival function is
\begin{align}
S(x)
=
\frac{x^{1-\alpha}-L_{\max}^{1-\alpha}}
{L_{\min}^{1-\alpha}-L_{\max}^{1-\alpha}},
\qquad
L_{\min}\leq x\leq L_{\max},
\end{align}
with $S(x)=1$ for $x<L_{\min}$ and $S(x)=0$ for $x>L_{\max}$.
Since $\alpha>1$, in the intermediate range
$L_{\min}\leq x\ll L_{\max}$, the survival function decays as
$S(x)\propto x^{-(\alpha-1)}$.

Under steady sliding at velocity $V$, Eq.~\eqref{eq:Pt_sol}
reduces to $P_{\mathrm{ss}}(\theta)
= VS(V\theta)/\langle L\rangle_{\mathrm{c}}$.
Substituting $x=V\theta$ into the asymptotic form of $S(x)$
yields $P_{\mathrm{ss}}(\theta)\propto\theta^{-(\alpha-1)}$
for $L_{\min}/V\leq \theta\ll L_{\max}/V$.
After the velocity step, the new contacts ($\theta<t$) in
Eq.~\eqref{eq:Pt_step} take the steady-state form
$P_t(\theta)\propto S(V_2\theta)$ and thus follow the same
power law $\theta^{-(\alpha-1)}$ in the intermediate range 
as shown in Fig.~\ref{fig3}.

\begin{figure}[!tbp]
    \centering
    \includegraphics[width=0.85\columnwidth]{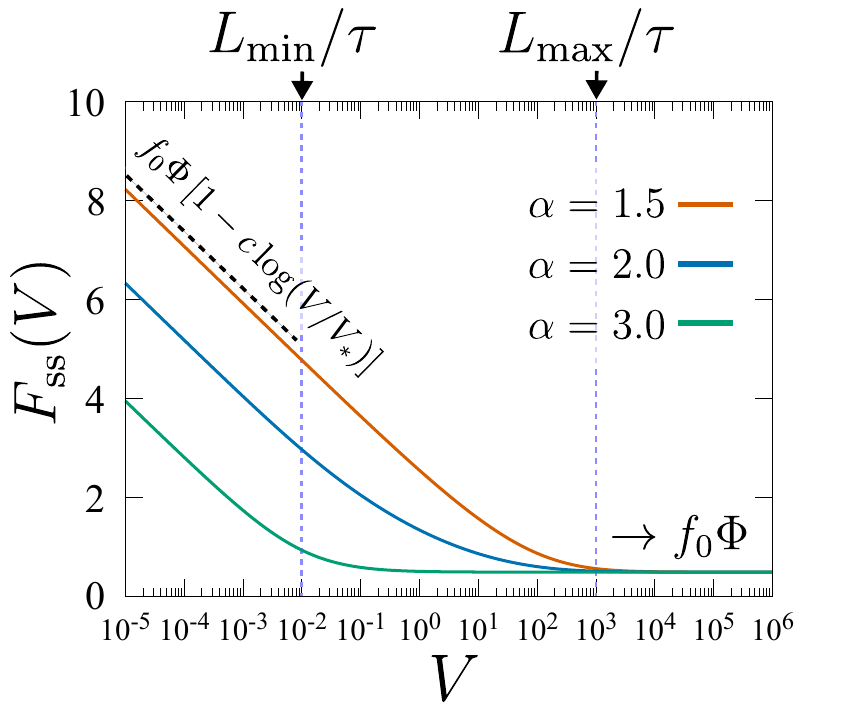}
    \caption{
        Velocity dependence of the steady-state friction force
        $F_{\mathrm{ss}}(V)$ for $\alpha=1.5,\,2.0,\,3.0$.
        For all $\alpha$, $F_{\mathrm{ss}}(V)$ exhibits monotonic
        velocity weakening.
        In the low-velocity regime $V\tau\ll L_{\min}$, it decreases
        linearly in $\log V$ with the $\alpha$-independent slope
        $-f_0\Phi c$ [Eq.~\eqref{eq:EM_Fss_lowvel}], whereas in the
        high-velocity regime $V\tau\gg L_{\max}$, it approaches
        $f_0\Phi$, the value in the absence of aging ($c=0$),
        as $V^{-1}$ [Eq.~\eqref{eq:EM_Fss_highvel}].
        The other parameters are the same as in Fig.~\ref{fig2}.
    }
    \label{fig5}
\end{figure}

\bigskip
\sectionprl{Appendix B: Velocity dependence of $F_{\mathrm{ss}}(V)$}
\renewcommand{\theequation}{B\arabic{equation}}
\renewcommand{\theHequation}{B.\arabic{equation}}
\setcounter{equation}{0}
The steady-state friction force $F_{\mathrm{ss}}(V)$ [Eq.~\eqref{eq:Fss}]
can be rewritten as
\begin{align}\label{eq:EM_Fss_H}
    F_{\mathrm{ss}}(V)=f_0\Phi
    \left[
        1+\frac{c V\tau}{\langle L \rangle_{\mathrm{c}}}
        \left\langle H\!\left( \frac{L}{V\tau} \right)
        \right\rangle_{\mathrm{c}}
    \right],
\end{align}
where $H(y)\equiv (1+y)\log(1+y)-y$.
As shown in Fig.~\ref{fig5}, $F_{\mathrm{ss}}(V)$ is a monotonically
decreasing function of $V$ for any $\alpha$. The steady-state friction
is thus velocity weakening.
This is because faster sliding shortens the contact lifetime,
so that contacts rupture before they are strengthened by aging.

In the low-velocity regime $\epsilon \equiv V\tau/L_{\min}\ll 1$,
asymptotic evaluation of Eq.~\eqref{eq:EM_Fss_H} yields
\begin{equation}\label{eq:EM_Fss_lowvel}
    F_{\mathrm{ss}}(V) = f_0\Phi
    \left[
        1 - c\log\left(\frac{V}{V_*}\right)
        + O\!\left(\epsilon\log\epsilon^{-1}\right)
    \right],
\end{equation}
where
$V_* \equiv (L_0/\tau)
\exp\!\left[\langle L\log(L/L_0)\rangle_{\mathrm{c}}
/\langle L\rangle_{\mathrm{c}}-1\right]$
is an intrinsic reference velocity determined by the
contact-length distribution and is independent of the arbitrary
reference length $L_0$.
Equation~\eqref{eq:EM_Fss_lowvel} shows that, at low velocities,
$F_{\mathrm{ss}}(V)$ decreases linearly in $\log V$ with the
$\alpha$-independent slope $dF_{\mathrm{ss}}/d\log V=-f_0\Phi c$.
The distribution enters only through $V_*$, which sets the
horizontal position of the curve (Fig.~\ref{fig5}).

In the high-velocity regime $V\tau\gg L_{\max}$, on the other hand,
we obtain
\begin{align}\label{eq:EM_Fss_highvel}
    F_{\mathrm{ss}}(V)= f_0\Phi
    \left[
        1 + \frac{c\langle L^2 \rangle_{\mathrm{c}}}
        {2\langle L \rangle_{\mathrm{c}}V\tau}
        \left\{1+O\!\left(\frac{L_{\max}}{V\tau}\right)\right\}
    \right].
\end{align}
In this regime, the contact lifetime is at most
$L_{\max}/V\ll\tau$, so that aging barely proceeds before rupture.
Consequently, $F_{\mathrm{ss}}(V)$ approaches the aging-free value
$f_0\Phi$.

\bigskip
\sectionprl{Appendix C: Relaxation of initial and final regimes}
\renewcommand{\theequation}{C\arabic{equation}}
\renewcommand{\theHequation}{C.\arabic{equation}}
\setcounter{equation}{0}
In the initial regime $u\ll\min\{1,\xi_{\min}\}$, the time elapsed
after the velocity step is short compared with both the aging
time $\tau$ and the renewal time $L_{\min}/V_2$ of the shortest
contacts, so that the replacement of the contact population has
barely proceeded. Accordingly, for fixed $r\neq1$,
\begin{equation}\label{eq:EM_R_initial}
    R(u) = 1 + O\!\left(\frac{u}{\min\{1,\xi_{\min}\}}\right),
\end{equation}
and the relaxation function exhibits the initial plateau seen in
Fig.~\ref{fig4}.

Near the end of the relaxation, $u\to\xi_{\max}$, almost all the
contacts formed before the step have already been ruptured, and only
those with $\xi'$ close to the upper cutoff $\xi_{\max}$ survive.
Writing $\Delta\equiv\xi_{\max}-u$ and setting $\xi'=u+y$ with
$0<y<\Delta$ in Eq.~\eqref{eq:N_def},
for $\Delta\ll\xi_{\max}$ the survival function vanishes linearly
at the cutoff, $s(u+y)\propto\Delta-y$ (Appendix~A), while the
aging difference is linear in $y$,
$z(u+r^{-1}y)-z(u+y)\simeq(r^{-1}-1)\,z'(u)\,y\propto y$.
Consequently,
$\mathcal{N}(u)\propto\int_0^\Delta(\Delta-y)\,y\,dy\propto\Delta^3$.
Thus, irrespective of $\alpha$, the contribution of the remaining
contacts vanishes cubically:
\begin{equation}\label{eq:EM_R_final}
    R(u)\propto(\xi_{\max}-u)^3,
\end{equation}
and the relaxation completes at $u=\xi_{\max}$.

\begin{figure}[!tbp]
    \centering
    \includegraphics[width=0.9\columnwidth]{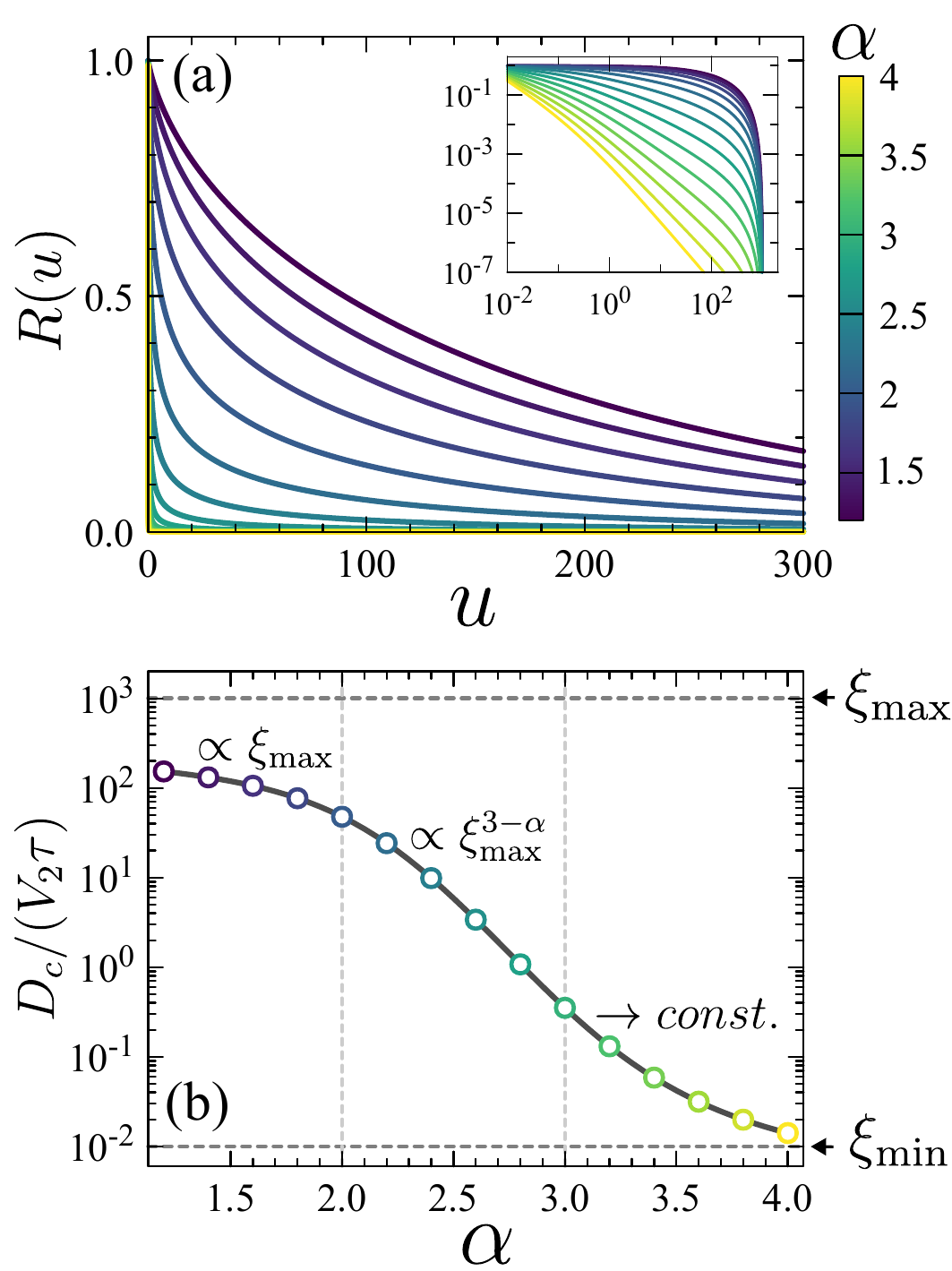}
    \caption{
        (a) Relaxation function $R(u)$ for $\alpha=1.2$--$4.0$
        (color scale), shown on linear axes with the same data on
        logarithmic axes in the inset.
        (b) $D_c/(V_2\tau)$ versus $\alpha$.
        It scales as $\xi_{\max}$ for $\alpha<2$ and as
        $\xi_{\max}^{\,3-\alpha}$ for $2<\alpha<3$, and saturates at a
        value set by $\xi_{\min}$ for $\alpha>3$ \cite{SM}.
        Horizontal dotted lines indicate $\xi_{\min}$ and 
        $\xi_{\max}$. Vertical dotted lines mark the regime 
        boundaries $\alpha=2$ and $3$.
        The parameters are the same as in Fig.~\ref{fig2}.
    }
    \label{fig6}
\end{figure}

\bigskip
\sectionprl{Appendix D: Slowing down of the relaxation for 
smaller $\alpha$}
\renewcommand{\theequation}{D\arabic{equation}}
\renewcommand{\theHequation}{D.\arabic{equation}}
\setcounter{equation}{0}
Figure~\ref{fig6}(a) shows the relaxation function $R(u)$ for
$\alpha\in [1.2,4.0]$.
At every $u$, $R(u)$ decreases monotonically with $\alpha$,
so that the relaxation slows down systematically as $\alpha$ decreases.
This is because a smaller $\alpha$ shifts the weight of the
contact-length distribution toward long contacts near the upper
cutoff $L_{\max}$, whose slow renewal sets the relaxation rate,
consistent with the asymptotic forms in Eq.~\eqref{eq:relax_scaling}.

To quantify this trend, we define the characteristic relaxation slip
distance
\begin{equation}\label{eq:EM_Dc_def}
    D_c \equiv \int_0^\infty R\left(\frac{X}{V_2\tau}\right)\,dX
    = V_2\tau\int_0^\infty R(u)\,du,
\end{equation}
which reduces to the decay length for an exponential relaxation
and plays the role of the characteristic slip distance of the
rate-and-state laws \cite{Ruina1983,Marone1998}.

As shown in Fig.~\ref{fig6}(b), $D_c/(V_2\tau)$ decreases from a value 
of order $\xi_{\max}$ to one of order $\xi_{\min}$ as $\alpha$ increases 
from $1.2$ to $4.0$.
Evaluating Eq.~\eqref{eq:EM_Dc_def}, we find that for $\alpha<3$,
$D_c$ is governed by the longest contacts near the upper cutoff,
giving $D_c/(V_2\tau)\sim\xi_{\max}$ for $1<\alpha<2$ and
$\xi_{\max}^{\,3-\alpha}$ for $2<\alpha<3$.
For $\alpha>3$, $D_c$ is instead governed by the shortest contacts
near the lower cutoff, so that it becomes independent of $L_{\max}$
and saturates at a value set by the microscopic scales $L_{\min}$
and $V_2 \tau$.
See Supplemental Material \cite{SM} for the derivation.

\renewcommand{\theHequation}{S.\arabic{equation}}




\clearpage

\setcounter{page}{1}
\renewcommand{\thepage}{S\arabic{page}}

\onecolumngrid

\begin{center}
{\bfseries\large
Supplemental Material for\\
``Relaxation of sliding friction from a statistical model of aging contacts''
}

\vspace{1em}

Ryudo Suzuki

\vspace{0.5em}

{\itshape
Department of Physics, Kyoto University, Kyoto 606-8502, Japan
}
\end{center}

\vspace{1.5em}

\renewcommand{\theequation}{S\arabic{equation}}
\setcounter{equation}{0}

\renewcommand{\thefigure}{S\arabic{figure}}
\setcounter{figure}{0}

This Supplemental Material collects the derivations omitted from the
main text and the End Matter.
In Sec.~\hyperref[sec:S1]{S1}, we solve the master equation
\eqref{eq:master_eq} by the method of characteristics and derive
Eq.~\eqref{eq:Pt_sol} of the main text.
In Sec.~\hyperref[sec:S2]{S2}, we evaluate the relaxation function
in the intermediate-time regime and derive Eq.~\eqref{eq:R_asymptotic}.
In Sec.~\hyperref[sec:S3]{S3}, we determine the asymptotic forms of
$I_\alpha(\Lambda)$ and $\mathcal{N}(0)$, which yield the relaxation
laws in Eq.~\eqref{eq:relax_scaling}.
In Sec.~\hyperref[sec:S4]{S4}, we derive the scaling of the
characteristic slip distance $D_c$ with the upper cutoff
$\xi_{\max}$.

\section{S1. Derivation of Eq.~(9)}\label{sec:S1}
For $\theta>0$, the influx term vanishes, and Eq.~\eqref{eq:master_eq}
reduces to
\begin{align}\label{eq:master_eq_bulk}
    \frac{\partial P_t(\theta)}{\partial t}
    + \frac{\partial P_t(\theta)}{\partial \theta}
    = -V(t)\lambda(x(t,\theta)) P_t(\theta).
\end{align}
We solve this equation under the boundary condition $P_t(0)=J(t)$.

The left-hand side of Eq.~\eqref{eq:master_eq_bulk} is the derivative
along the characteristic $(t',\theta')=(t-\theta+s,s)$.
We thus fix arbitrary $t$ and $\theta\,(>0)$ and introduce the
function along the characteristic line
\begin{align}
    p(s)\equiv P_{t-\theta+s}(s),\qquad 0\leq s\leq\theta.
\end{align}
At $s=0$, the boundary condition gives $p(0)=J(t-\theta)$, and at
$s=\theta$, $p(\theta)=P_t(\theta)$.
From Eq.~\eqref{eq:master_eq_bulk}, $p(s)$ obeys the ordinary
differential equation
\begin{align}
    \frac{d p(s)}{ds}
    = -V(t-\theta+s)\,\lambda(x(t-\theta+s,s))\,p(s).
\end{align}
Separating variables and integrating from $s=0$ to $s=\theta$, 
we obtain
\begin{equation}\label{eq:characteristic_int}
    \begin{aligned}
        \log\frac{p(\theta)}{p(0)}
        &=
        -\int_0^\theta
        V(t-\theta+s)\,
        \lambda\!\left(x(t-\theta+s,s)\right)\,ds
        \\
        &=
        -\int_0^{x(t,\theta)}\lambda(x)\,dx
        \\
        &=
        \log S\!\left(x(t,\theta)\right).
    \end{aligned}
\end{equation}
In the second line, we used the change of variables
$x=x(t-\theta+s,s)=\int_{t-\theta}^{t-\theta+s}V(s')ds'$
($dx=V(t-\theta+s)\,ds$).
In the third line, we used
$\lambda(x)=-\frac{d}{dx}\log S(x)$ and $S(0)=1$.
Since $p(0)=J(t-\theta)$ and $p(\theta)=P_t(\theta)$, 
we obtain
\begin{align}\label{eq:Pt_sol_J}
    P_t(\theta)=J(t-\theta)\,S(x(t,\theta)).
\end{align}
Thus, the influx $J(t-\theta)$ is weighted by the probability
$S(x(t,\theta))$ that a contact formed at time $t-\theta$ survives
after traveling the distance $x(t,\theta)$.

Next, we determine $J(t)$ self-consistently.
Substituting Eq.~\eqref{eq:Pt_sol_J} into Eq.~\eqref{eq:flux} and
using $\lambda(x)S(x)=\rho_{\mathrm{c}}(x)$ gives
\begin{align}\label{eq:self_consistent}
    J(t)=V(t) \int_0^\infty
    \rho_{\mathrm{c}}(x(t,\theta))\,J(t-\theta)\,d\theta.
\end{align}
A solution is
$J(t)=V(t)/\langle L\rangle_{\mathrm c}$.
Indeed, substituting it into the right-hand side and 
using $x=x(t,\theta)$, with $dx=V(t-\theta)\,d\theta$, gives
\begin{align}
    V(t)\int_0^\infty \rho_{\mathrm{c}}(x(t,\theta))\,
    \frac{V(t-\theta)}{\langle L\rangle_{\mathrm{c}}}\, d\theta
    = \frac{V(t)}{\langle L\rangle_{\mathrm{c}}}
    \int_0^\infty \rho_{\mathrm{c}}(x)\,dx
    =\frac{V(t)}{\langle L\rangle_{\mathrm{c}}}
    =J(t).
\end{align}
Substituting this result into Eq.~\eqref{eq:Pt_sol_J} yields
Eq.~\eqref{eq:Pt_sol} in the main text.

\section{S2. Derivation of Eq.~(16)}\label{sec:S2}
We evaluate the numerator $\mathcal{N}(u)$ [Eq.~\eqref{eq:N_def}]
in the intermediate-time regime
\begin{align}\label{eq:scaling_window}
    \max\{1,\xi_{\min}\}\ll u \ll \xi_{\max}.
\end{align}
Changing variables from $\xi'$ to $x=\xi'/u$ gives
\begin{align}\label{eq:N_x}
    \mathcal{N}(u)
    = u\int_1^\infty dx\, s(ux)
    \left[
        z\!\left(u\Bigl(1+\tfrac{x-1}{r}\Bigr)\right) -z(ux)
    \right].
\end{align}
This expression does not depend on the specific functional forms of
$Z(\theta)$ and $\rho_{\mathrm c}(L)$.

For the logarithmic aging function
$z(y)=1+c\log(1+y)$ [Eq.~\eqref{eq:aging}], and for $x\geq 1$ and $u\gg 1$,
\begin{align}\label{eq:z_expansion}
    z\!\left(u\Bigl(1+\tfrac{x-1}{r}\Bigr)\right) -z(ux)
    &=c \log\!\left[
    \frac{1+u\bigl(1+\tfrac{x-1}{r}\bigr)}{1+ux}
    \right]\notag\\
    &=c g(x)+O\!\left((ux)^{-1}\right),
\end{align}
where $g(x)=\log\left[(1+(x-1)/r)/x\right]$.
Here, we factored out the leading terms proportional to $u$ in the
numerator and denominator, using $ux\geq u\gg1$.

The survival function corresponding to the truncated power-law
distribution in Eq.~\eqref{eq:size_dist} is
\begin{align}\label{eq:survival_function}
    s(y)=
    \begin{cases}
        1, &y\leq \xi_{\min},\\
        (y^{1-\alpha}-\xi_{\max}^{1-\alpha})/A, &\xi_{\min}\leq y\leq \xi_{\max},\\
        0, &y\geq \xi_{\max}.
    \end{cases}
\end{align}
Since $s(y)=0$ for $y\geq \xi_{\max}$
[Eq.~\eqref{eq:survival_function}], defining
$\Lambda\equiv \xi_{\max}/u$ restricts the integral in
Eq.~\eqref{eq:N_x} to $1\leq x \leq \Lambda$.
Furthermore, in the intermediate-time regime, 
$ux\geq u\gg \xi_{\min}$, and hence
\begin{align}
    s(ux)
    =
    \frac{u^{1-\alpha}}{A}
    \left(x^{1-\alpha}-\Lambda^{1-\alpha}\right).
\end{align}
Substituting this expression and Eq.~\eqref{eq:z_expansion}
into Eq.~\eqref{eq:N_x}, and dividing by $\mathcal{N}(0)$,
which is independent of $u$, we obtain
\begin{equation}\label{eq:R_intermediate}
    R(u)=\frac{\mathcal{N}(u)}{\mathcal{N}(0)}
    = \frac{c\,u^{2-\alpha}}{A\,\mathcal{N}(0)}
    \left[\,I_\alpha(\Lambda)
    +O\!\left(u^{-1}\right)\right].
\end{equation}
Here, the $O(u^{-1})$ term bounds the contribution of the error in
Eq.~\eqref{eq:z_expansion}: since
$(x^{1-\alpha}-\Lambda^{1-\alpha})\,x^{-1}\leq x^{-\alpha}$ is
integrable for $\alpha>1$, that contribution is at most
$O(u^{-1})\int_1^\infty x^{-\alpha}\,dx=O(u^{-1})$, uniformly in
$\Lambda$.
As shown in Sec.~S3, for fixed $r\neq1$ and $\Lambda\gg1$,
$I_\alpha(\Lambda)$ either grows with $\Lambda$ or approaches a
nonzero constant.
Therefore, the correction is of relative order at most
$O(u^{-1})$, yielding Eq.~\eqref{eq:R_asymptotic} in the
main text.

\section{S3. Derivation of Eq.~(17)}\label{sec:S3}
To evaluate Eq.~\eqref{eq:R_intermediate}, we determine the asymptotic 
scaling of $I_\alpha(\Lambda)$ and of $\mathcal{N}(0)$.

\subsection{Asymptotics of $I_\alpha(\Lambda)$}
We decompose the logarithmic factor $g(x)$ into a constant term 
independent of $x$ and a term that vanishes as $x\to \infty$:
\begin{align}\label{eq:g_decomp}
    g(x)
    =\beta_r+\log\!\left(1+\frac{r-1}{x}\right),
    \qquad \beta_r\equiv\log\!\left(\frac{1}{r}\right).
\end{align}
Accordingly, $I_\alpha$ is decomposed as
\begin{equation}\label{eq:Ialpha_split}
    I_\alpha(\Lambda)
    = \beta_r\, K_\alpha(\Lambda)
    +\int_1^\Lambda dx\,
    \bigl(x^{1-\alpha}-\Lambda^{1-\alpha}\bigr)
    \log\!\left(1+\frac{r-1}{x}\right),
    \qquad
    K_\alpha(\Lambda)
    \equiv\int_1^\Lambda dx\,
    \bigl(x^{1-\alpha}-\Lambda^{1-\alpha}\bigr).
\end{equation}
The integral $K_\alpha(\Lambda)$ is evaluated exactly as
\begin{align}\label{eq:first_term}
    K_\alpha(\Lambda)
    =
    \begin{cases}
        \dfrac{\alpha-1}{2-\alpha}\,\Lambda^{2-\alpha}
        -\dfrac{1}{2-\alpha}+\Lambda^{1-\alpha},
        & \alpha\neq2,\\[8pt]
        \log\Lambda-1+\Lambda^{-1},
        & \alpha=2,
    \end{cases}
\end{align}
which scales, for $\Lambda\gg1$, as $\Lambda^{2-\alpha}$ for
$1<\alpha<2$ and as $\log\Lambda$ for $\alpha=2$, and approaches
$1/(\alpha-2)$ for $\alpha>2$.
In the second term of Eq.~\eqref{eq:Ialpha_split}, since
$\log(1+(r-1)/x)=O(x^{-1})$ as $x\to\infty$, the integrand decays as
$O(x^{-\alpha})$. The integral therefore converges as
$\Lambda\to\infty$ for $\alpha>1$ and contributes $O(1)$.
Hence, for $\Lambda\gg1$,
\begin{align}\label{eq:Ialpha_asymptotic}
    I_\alpha(\Lambda)
    =
    \begin{cases}
        \beta_r\dfrac{\alpha-1}{2-\alpha}\Lambda^{2-\alpha}
        +O(1),
        & 1<\alpha<2,\\
        \beta_r\log\Lambda+O(1),
        & \alpha=2,\\
        I_\alpha^{\infty}(r)+O(\Lambda^{2-\alpha}),
        & \alpha>2,
    \end{cases}
    \qquad
    I_\alpha^{\infty}(r)\equiv\int_1^\infty dx\, x^{1-\alpha}\, g(x),
\end{align}
where the integral defining $I_\alpha^{\infty}(r)$ converges for 
$\alpha>2$ because $g(x)\to \beta_r$ as $x\to\infty$.

\subsection{Evaluation of $\mathcal{N}(0)$}

Setting $u=0$ in Eq.~\eqref{eq:N_def} gives
\begin{align}\label{eq:SM_N0}
    \mathcal{N}(0)
    =\int_0^\infty d\xi\, s(\xi)
    \left[z(\xi/r)-z(\xi)\right].
\end{align}
For the aging function Eq.~\eqref{eq:aging},
$z(\xi/r)-z(\xi)=c\log[(1+\xi/r)/(1+\xi)]=c\,\psi'(\xi)$, where
\begin{align}\label{eq:psi_def}
    \psi(\xi)\equiv r\,H\!\left(\frac{\xi}{r}\right)-H(\xi),
\end{align}
with $H(y)= (1+y)\log(1+y)-y$, defined in Eq.~\eqref{eq:EM_Fss_H}.
Here we used $H'(y)=\log(1+y)$ and $\psi(0)=0$.
Integrating Eq.~\eqref{eq:SM_N0} by parts, we obtain
\begin{align}\label{eq:N0_exact}
    \mathcal{N}(0)
    =c\left\langle \psi(\xi)\right\rangle_{\mathrm{c}}
    =\frac{c}{A} J_{\alpha}(\xi_{\min},\xi_{\max}),
    \qquad J_{\alpha}(a,b)\equiv 
    (\alpha-1) \int_{a}^{b} d\xi\,\xi^{-\alpha}\psi(\xi).
\end{align}
Here $\xi\equiv L/(V_2\tau)$, and for any function $f$ we write
$\langle f(\xi)\rangle_{\mathrm{c}}\equiv
\langle f(L/(V_2\tau))\rangle_{\mathrm{c}}$.
We now derive the asymptotic form of $J_\alpha(a,b)$ for
$b\gg\max\{a,1,r\}$.
Noting that, as in Eq.~\eqref{eq:g_decomp}, the derivative
decomposes as
$\psi'(\xi)=\beta_r+\log[(\xi+r)/(\xi+1)]$,
integration by parts gives
\begin{align}\label{eq:J_ibp}
    J_{\alpha}(a,b)
    =
    \begin{cases}
        -b^{1-\alpha}\psi(b)+a^{1-\alpha}\psi(a)
        +\dfrac{\beta_r}{2-\alpha}
        \left(b^{2-\alpha}-a^{2-\alpha}\right)
        +\displaystyle\int_a^b d\xi\,
        \xi^{1-\alpha}\log\frac{\xi+r}{\xi+1},
        & \alpha\neq2,
        \\[1.2em]
        -\dfrac{\psi(b)}{b}
        +\dfrac{\psi(a)}{a}
        +\beta_r\log\dfrac{b}{a}
        +\displaystyle\int_a^b \frac{d\xi}{\xi}
        \log\frac{\xi+r}{\xi+1},
        & \alpha=2.
    \end{cases}
\end{align}
Here, $\psi$ behaves for $\xi\gg\{1,r\}$ as
\begin{align}\label{eq:psi_asym}
    \psi(\xi)=\beta_r\,\xi+(r-1)\log\xi+O(1).
\end{align}
Moreover, in the last integrals of Eqs.~\eqref{eq:J_ibp}, 
the integrand decays as $O(\xi^{-\alpha})$
because $\log[(\xi+r)/(\xi+1)]=O(\xi^{-1})$. These integrals
therefore converge as $b\to\infty$ for $\alpha>1$ and contribute
only $O(1)$.

For $1<\alpha<2$, Eq.~\eqref{eq:psi_asym} gives
$-b^{1-\alpha}\psi(b)
=-\beta_r b^{2-\alpha}+O(b^{1-\alpha}\log b)$,
which combines with the term
$\beta_r b^{2-\alpha}/(2-\alpha)$ in Eq.~\eqref{eq:J_ibp}
to yield the coefficient $\beta_r(\alpha-1)/(2-\alpha)$.
The $a$-dependent terms are $O(1)$, so
\begin{align}\label{eq:J_alpha_less_2}
    J_{\alpha}(a,b)=\beta_r \frac{\alpha-1}{2-\alpha}\,b^{2-\alpha}+O(1).
\end{align}

For $\alpha=2$, Eq.~\eqref{eq:psi_asym} gives
$-\psi(b)/b=-\beta_r+O(b^{-1}\log b)$, which remains $O(1)$.
The only divergent term is $\beta_r\log b$, and hence
\begin{align}\label{eq:J_alpha2}
    J_2(a,b)=\beta_r \log b+O(1).
\end{align}

For $\alpha>2$, Eq.~\eqref{eq:psi_asym} shows that the integrand
decays as $\xi^{1-\alpha}$, so the integral itself converges at
the upper limit:
\begin{align}\label{eq:J_alpha_large_2}
    J_\alpha(a,b)=
    J_\alpha^\infty(r,\xi_{\min})+O(b^{2-\alpha}),\qquad 
    J_\alpha^\infty(r,\xi_{\min})\equiv
    (\alpha-1)\int_{\xi_{\min}}^\infty d\xi \xi^{-\alpha}\psi(\xi),
\end{align}
where we used 
$\int_b^\infty d\xi\, \xi^{-\alpha}\psi(\xi)=O(b^{2-\alpha})$
for the truncated tail.

Substituting
Eqs.~\eqref{eq:J_alpha_less_2}--\eqref{eq:J_alpha_large_2} with
$a=\xi_{\min}$ and $b=\xi_{\max}$
into Eq.~\eqref{eq:N0_exact}, we obtain, for
$\xi_{\max}\gg\{1,r,\xi_{\min}\}$,
\begin{align}\label{eq:S4_N0_scaling}
    A\mathcal{N}(0)
    = c
    \begin{cases}
        \beta_r\frac{\alpha-1}{2-\alpha}\xi_{\max}^{2-\alpha}+O(1),&1<\alpha<2,\\
        \beta_r \log\xi_{\max}+O(1),&\alpha=2,\\
        J_\alpha^\infty(r,\xi_{\min})+O(\xi_{\max}^{2-\alpha}),&\alpha>2.
    \end{cases}
\end{align}

\subsection{Asymptotic form of $R(u)$}
In the intermediate-time regime \eqref{eq:scaling_window},
$\Lambda=\xi_{\max}/u\gg1$.
Substituting Eqs.~\eqref{eq:Ialpha_asymptotic} and 
\eqref{eq:S4_N0_scaling} into Eq.~\eqref{eq:R_intermediate}, 
we obtain
\begin{align}\label{eq:R_final}
    R(u)=
    \begin{cases}
        1+O\!\left((u/\xi_{\max})^{2-\alpha}\right)
        +O\!\left(\xi_{\max}^{\alpha-2}\right),
        & 1<\alpha<2,\\[4pt]
        1-\dfrac{\log u}{\log\xi_{\max}}
        +O\!\left(\dfrac{1}{\log\xi_{\max}}\right),
        & \alpha=2,\\[10pt]
        \mathcal{C}_\alpha \,u^{-(\alpha-2)}
        \left[1+O(u^{-1})
        +O\!\left(\left(\dfrac{u}{\xi_{\max}}\right)^{\alpha-2}\right)
        \right],
        & \alpha>2,
    \end{cases}
\end{align}
where $\mathcal{C}_\alpha \equiv I_\alpha^\infty(r)/J_\alpha^\infty(r,\xi_{\min})$,
which is Eq.~\eqref{eq:relax_scaling} in the main text.

\section{S4. Scaling of $D_c$}\label{sec:S4}
 
We determine the scaling of the characteristic relaxation slip 
distance $D_c$ with the upper cutoff $\xi_{\max}$.
From Eqs.~\eqref{eq:EM_Dc_def} and \eqref{eq:R_express},
\begin{align}\label{eq:Dc_I}
    \frac{D_c}{V_2\tau}
    =
    \frac{\mathcal I}{\mathcal N(0)},
    \qquad
    \mathcal I\equiv
    \int_0^\infty \mathcal N(u)\,du .
\end{align}
Substituting Eq.~\eqref{eq:N_def} and exchanging the order of
integration gives
\begin{align}\label{eq:Dc_W}
    \mathcal I
    =
    \int_0^\infty d\xi\,s(\xi)\,W(\xi),\qquad
    W(\xi)
    \equiv
    \int_0^\xi du\,
    \left[
        z\!\left(u+r^{-1}(\xi-u)\right)-z(\xi)
    \right].
\end{align}
The $u$ integral gives
\begin{align}\label{eq:W_exact}
    W(\xi)
    =
    c\left[
        \frac{\psi(\xi)}{1-r}
        +\log(1+\xi)-\xi
    \right],
\end{align}
where $\psi$ is defined in Eq.~\eqref{eq:psi_def}.
 
Let $G'(\xi)=W(\xi)$ with $G(0)=0$.
Integration by parts then gives
$\mathcal I=\langle G(\xi)\rangle_{\mathrm c}$.
Using Eq.~\eqref{eq:psi_asym}, we find, for $\xi\gg\max\{1,r\}$,
\begin{align}\label{eq:G_asym}
    G(\xi)
    =
    c\gamma_r\,\xi^2+O(\xi),
    \qquad
    \gamma_r
    \equiv
    \frac{r-1-\log r}{2(1-r)} .
\end{align}
Therefore, averaging Eq.~\eqref{eq:G_asym} over $\rho_{\mathrm{c}}$
and applying Eq.~\eqref{eq:psi_asym} likewise to
$\mathcal N(0)=c\langle\psi(\xi)\rangle_{\mathrm c}$
[Eq.~\eqref{eq:N0_exact}], we obtain
\begin{align}\label{eq:I_N0}
    \mathcal I
    =
    c\gamma_r\langle\xi^2\rangle_{\mathrm c}
    +O\!\left(\langle\xi\rangle_{\mathrm c}\right),
    \qquad
    \mathcal N(0)
    =
    c\beta_r\langle \xi\rangle_{\mathrm{c}}+O(1);
\end{align}
the scalings of $\mathcal I$ and $\mathcal N(0)$ are thus governed
by the second and first moments of the contact length, respectively.
For fixed $\xi_{\min}$ and $\xi_{\max}\to\infty$, the truncated
power law in Eq.~\eqref{eq:size_dist} gives
\begin{align}\label{eq:moment_scaling}
    \langle \xi^n\rangle_{\mathrm{c}}
    \sim
    \begin{cases}
        \xi_{\max}^{\,n+1-\alpha},
        & \alpha<n+1,\\
        \log\xi_{\max},
        & \alpha=n+1,\\
        O(1),
        & \alpha>n+1.
    \end{cases}
\end{align}
Using Eq.~\eqref{eq:moment_scaling} with $n=1,2$ in
Eqs.~\eqref{eq:Dc_I} and \eqref{eq:I_N0}, we obtain, for fixed
$\xi_{\min}$ and $\xi_{\max}\gg\{1,r,\xi_{\min}\}$,
\begin{align}\label{eq:Dc_scaling}
    \frac{D_c}{V_2\tau}
    \sim
    \begin{cases}
        \xi_{\max},
        & 1<\alpha<2,\\[2pt]
        \xi_{\max}/\log\xi_{\max},
        & \alpha=2,\\[2pt]
        \xi_{\max}^{\,3-\alpha},
        & 2<\alpha<3,\\[2pt]
        \log\xi_{\max},
        & \alpha=3,\\[2pt]
        O(1),
        & \alpha>3.
    \end{cases}
\end{align}
 
For $\alpha>3$, both $\mathcal I$ and $\mathcal N(0)$ converge as
$\xi_{\max}\to\infty$: $D_c$ is then independent of $L_{\max}$ and
is set by the microscopic scales $L_{\min}$ and $V_2\tau$.

\end{document}